\documentclass[aps,pra,longbibliography,11pt,reprint,floatfix,superscriptaddress,urlname]{revtex4-2}
\usepackage[utf8]{inputenc}
\usepackage{amsmath}
\usepackage{amsfonts}
\usepackage{color}
\usepackage{graphicx}
\usepackage{comment}
\usepackage{caption}
\usepackage{subcaption}
\usepackage[normalem]{ulem}
\usepackage{multirow}
\usepackage{hyperref}
\usepackage[capitalise]{cleveref}
\usepackage{lineno} 

\begin{document}

\title{On the Discretization Error of the Discrete Generalized Quantum Master Equation}

\author{Ruojing Peng}
\affiliation{Department of Chemistry and Chemical Biology, Harvard University, Cambridge, MA 02138, USA}

\author{Lachlan P. Lindoy}
\affiliation{National Physical Laboratory, Teddington, TW11 0LW, United Kingdom}

\author{Joonho Lee}
\email{joonholee@g.harvard.edu}
\affiliation{Department of Chemistry and Chemical Biology, Harvard University, Cambridge, MA 02138, USA}

\begin{abstract}
The transfer tensor method (TTM) [Cerrillo and Cao, Phys. Rev. Lett. 2014, 112, 110401] can be considered a discrete-time formulation of the Nakajima-Zwanzig quantum master equation (NZ-QME) for modeling non-Markovian quantum dynamics. A recent paper [Makri, J. Chem. Theory Comput. 2025, 21, 5037] raised concerns regarding the consistency of the TTM discretization, particularly a spurious term at the initial time $t=0$. This Communication presents a detailed analysis of the discretization structure of TTM, clarifying the origin of the initial-time correction and establishing a consistent relationship between the TTM discrete-time memory kernel $K_N$, and the continuous-time NZ-QME kernel $\mathcal{K}(N\Delta t)$. This relationship is validated numerically using the spin-boson model, demonstrating convergence of reconstructed memory kernels and accurate dynamical evolution as $\Delta t \to 0$. While TTM provides a consistent discretization, we note that alternative schemes are also viable, such as the midpoint derivative/midpoint integral scheme proposed in Makri's work. 
The relative performance of various schemes for either computing accurate $\mathcal{K}(N\Delta t)$ from exact dynamics, or obtaining accurate dynamics from exact $\mathcal{K}(N\Delta t)$, warrants further investigation.
\end{abstract}

\maketitle

\section{Introduction}
The Nakajima-Zwanzig quantum master equation (NZ-QME)~\cite{10.1143/PTP.20.948,10.1063/1.1731409} offers a formally exact approach for simulating non-Markovian open quantum dynamics, focusing only on quantities of the dimension of the systems. Of central importance in the NZ-QME formulism is the time-non-local memory kernel $\mathcal{K}(t)$, which accounts for the non-Markovian effect of environment, and is related to influence functional~\cite{ivander2024unified} of the path integral approach~\cite{FEYNMAN1963118,10.1063/1.469508,MAKAROV1994482,doi:10.1021/acs.jctc.0c00987,10.1063/1.5139473,10.1063/1.5084949,10.1063/1.4808108,10.1063/1.5058223,10.1063/5.0151748}. The NZ-QME memory kernel $\mathcal{K}(t)$ can be computed from various approximate schemes, such as Bloch-Redfield equations~\cite{PhysRev.70.460,redfield1957theory}, coupled integral equation~\cite{doi:10.1021/acs.jpcb.1c05719,10.1063/1.2218342,10.1063/1.1624830}, or from reduced dynamical channels on a dense time grid~\cite{10.1063/1.5055756,10.1063/1.5047446,doi:10.1021/acs.jpcb.9b08429}. A closely related alternative is the transfer tensor method (TTM)~\cite{PhysRevLett.112.110401} by Cerrillo and Cao, which computes discrete time memory kernels $K_n$ from dynamical channels on a coarse time grid. Being a highly efficient data-driven approach, TTM has demonstrated its efficiency and accuracy for a variety of systems ~\cite{10.21468/SciPostPhys.13.2.027,doi:10.1021/acs.jpclett.6b02389,PhysRevA.96.062122,10.1063/1.5009086}, and combines easily with quantum tomography  ~\cite{Pollock2018tomographically,PhysRevA.102.052206}.

In the original Refs.~\onlinecite{PhysRevLett.112.110401,Pollock2018tomographically}, the TTM discrete time memory kernel $K_n$ was simply related to the NZ-GME continuous time memory kernel $\mathcal{K}(n\Delta t)$ at each discrete time point (see, e.g., Eq. (5) in Ref.~\onlinecite{PhysRevLett.112.110401} and Eq. (12) in Ref.~\onlinecite{Pollock2018tomographically}). In addition, Ref.~\onlinecite{PhysRevLett.112.110401} numerically demonstrated that $K_n$ computed from TTM equation converges to $\mathcal{K}(n\Delta t)$ in the $\Delta t\to0$ limit for time points $n>0$ (Fig.4 in Ref.~\onlinecite{PhysRevLett.112.110401}). A recent paper~\cite{doi:10.1021/acs.jctc.5c00396} by Makri raised a concern on the relationship between $K_n$ and $\mathcal{K}(n\Delta t)$. In particular, $K_0$ is identified as a spurious term, which leads to wrong dynamics, while the simple identification of $K_n$ with $\mathcal{K}(n\Delta t)$ remains numerically valid for $n>0$. Furthermore, Ref.~\onlinecite{doi:10.1021/acs.jctc.5c00396} proposed an alternative discretization scheme for NZ-QME using exact $\mathcal{K}(n\Delta t)$. 

In this Communication, we present an analysis to clarify the apparent inconsistency in the discretized GQME. Section~\ref{sec:analyitical} provides a detailed analysis of the relation between $K_n$ from the TTM equations and $\mathcal{K}(n\Delta t)$ in the NZ-QME, with a consistent treatment of initial time, much of which was contained in the addendum to our recent work~\cite{ivander2024unified}.  Section~\ref{sec:numerics} provides numerical evidence for the relationship derived in Section~\ref{sec:analyitical}.

\section{Discretization error analysis}\label{sec:analyitical}
A detailed analysis of discretization error in the TTM equation is provided in Ref.~\onlinecite{ivander2024unified} (Addendum). Here, we provide a sketch of the analysis for completeness, emphasizing the subtleties associated with the initial time. 

The continuous-time homogeneous NZ equation~\cite{10.1143/PTP.20.948,10.1063/1.1731409,doi:10.1021/acs.jpcb.1c05719} for the system propagator $U(t)$ is
\begin{align}\label{eq:continuous_dot}
\dot{U}(t)=-iL_sU(t)+\int_0^td\tau\mathcal{K}(\tau)U(t-\tau)
\end{align}
where $L_s=[H_s,\cdot]$ and $U(t)$ is defined by $\rho(t)=U(t)\rho(0)$. For the subsequent analysis, it is useful to write its second-order derivative, as derived in Ref.~\citenum{10.1063/1.5047446}, 
\begin{align}\label{eq:continuous_ddot}
\ddot{U}(t)
&=(-iL_s)^2U(t)+\mathcal{K}(t)+\int_0^td\tau\mathcal{F}(\tau)U(t-\tau),
\end{align}
where 
\begin{align}\label{eq:F}
\mathcal{F}(t)=\{\mathcal{K}(t),-iL_s\}+\int_0^td\tau\mathcal{K}(\tau)\mathcal{K}(t-\tau),
\end{align}
and the third-order derivative 
\begin{align}\label{eq:continuous_dddot}
\dddot{U}(t)
&=\left[(-iL_s)^3+\{\mathcal{K}_0,-iL_s\}+\dot{\mathcal{K}}_0\right]U(t)\notag\\
&+\int_0^td\tau\mathcal{R}(t-\tau)U(\tau),
\end{align}
where
\begin{align}
\mathcal{R}(t)
&=\left[\left[(-iL_s)^2+\mathcal{K}_0\right]\mathcal{K}-iL_s\dot{\mathcal{K}}+\ddot{\mathcal{K}}\right](t).
\end{align}
We note that $\mathcal{F}(t)\sim\mathcal{O}(||\mathcal{K}(t)||)$ ($||\cdot||$ denotes Frobenius norm of the matrix) from Young's convolution inequality~\cite{doi:10.1098/rspa.1912.0086,Bogachev2006-uf}. 
Furthermore, $\dot{\mathcal{K}}(t)$, $\ddot{\mathcal{K}}(t)$, and hence $\mathcal{R}(t)$ are also $\mathcal{O}(||\mathcal{K}(t)||)$ from the projection of full system-and-bath evolution~\cite{10.1063/1.1624830,10.1063/1.2218342}.

We now consider the discrete-time version of \cref{eq:continuous_dot,eq:continuous_ddot,eq:continuous_dddot}. Following the same convention as our original manuscript (Ref.~\cite{ivander2024unified}), let $t=N\Delta t$, $L=I-i\Delta tL_s$, $\rho_N = \rho(N\Delta t)$, $U_N = U(N\Delta t)$. Furthermore, we denote by $\mathcal{K}_m=\mathcal{K}(m\Delta t)$ the continuous-time memory kernel in \cref{eq:continuous_dot} evaluated at discrete time steps, and denote by $K_m$ the discrete-time memory kernel defined by discrete time relations
\begin{align}\label{eq:nz_discrete}
U_{N+1}=LU_{N}+\Delta t^2\sum_{m=0}^{N}K_{m}U_{N-m},
\end{align}
from Ref.~\cite{ivander2024unified}. At this point, we also note that $\mathcal K_m$ may not be straightforwardly related to $K_m$ and could not be the same as $K_m$ in the limit of $\Delta t\rightarrow 0$. 
One must conduct a careful discretization analysis to examine this relationship and equivalence, which is the primary purpose of this Communication.
In the following analysis, we assume that $\{U_m\}$ are exact, which recovers the TTM~\cite{PhysRevLett.112.110401}. We demonstrate how the discretization error in the discrete NZ equation propagates into the relationship between $\mathcal K_m$ and $K_m$. 

We begin by considering the time derivatives of the system propagator at $t=0$ (i.e., $N=0$), 
\begin{align}
&\ddot{U}_0=(-iL_s)^2+\mathcal{K}_0,\\
&\dddot{U}_0=(-iL_s)^3+\{\mathcal{K}_0,-iL_s\}+\dot{\mathcal{K}}_0. \label{eq:discrete_dddot_0}
\end{align}
The expansion of  $U_1$ at $t=0$ follows,
\begin{align}\label{eq:U1}
U_1
&=I+\Delta t\dot{U_0}+\frac{\Delta t^2}{2}\ddot{U}_0+\frac{\Delta t^3}{6}\dddot{U}_0+\mathcal O(\Delta t^4).
\end{align}
We can use \cref{eq:U1,eq:nz_discrete} to obtain
\begin{align}\label{eq:K0}
K_0&=\frac{1}{2}\left((-iL_s)^2+\mathcal{K}_0\right)+\frac{\Delta t}{6}\dddot{U}_0+\mathcal O(\Delta t^2).
\end{align}
Note that \cref{eq:K0} is different from the simple assignment $K_0=\mathcal{K}_0$ in Ref.~\onlinecite{PhysRevLett.112.110401} (see Eq. (5) therein, and Eq. (12) in Ref.~\onlinecite{Pollock2018tomographically}). Here, the continuous time $\mathcal{K}_0$ is a well-defined {\it non-trivial} quantity, whose expression in terms of projectors is ~\cite{10.1063/1.1624830,10.1063/1.2218342} 
\begin{align}
\mathcal{K}_0=-\text{Tr}_\mathrm{bath}[L(I-P)L\rho_\mathrm{bath}],
\end{align}
where $L=[H,\cdot]$ is the Liouvillian of the full system and bath space, and $P(\cdot)=\rho_\mathrm{bath}\otimes\text{Tr}_\mathrm{bath}(\cdot)$ is the projector onto the system. In particular, the effect of $\mathcal{K}_0$ enters even in the propagation from $U_0$ to $U_1$, unlike hypothesized in Ref.~\onlinecite{doi:10.1021/acs.jctc.5c00396} (see, e.g., Eq. (12) and Eq. (18) therein). As worked out in the following, the ``convolution integral overestimation'' pointed out in Ref.~\onlinecite{doi:10.1021/acs.jctc.5c00396} can be considered as a way to correctly account for the $\mathcal{O}(\Delta t^2\ddot{U}_N)$ term in time-propagation from $U_N$ to $U_{N+1}$. 

For $N>0$, we discretize the time-convolution integral in \cref{eq:continuous_ddot,eq:continuous_dddot} with the right Riemann sum,
\begin{align}
\ddot{U}_N
&=(-iL_s)^2U_N+\mathcal{K}_N+\Delta t\sum_{m=1}^{N}\mathcal{F}_{m}U_{N-m}\notag\\&
+\Delta t^2\sum_{m=1}^{N}D^\mathcal{F}_{m},\label{eq:ddot_discrete}\\
\dddot{U}_N
&=\dddot{U}_0U_N+\Delta t\sum_{m=1}^{N}\mathcal{R}_{m}U_{N-m}+
\Delta t^2\sum_{m=1}^{N}D_{m}^{\mathcal{R}},\label{eq:dddot_discrete}
\end{align}
where 
\begin{align}
D_{m}^X&\sim\frac{d}{d\tau}\left[X(\tau)U(N\Delta t-\tau)\right]_{\tau=m\Delta t}
\end{align}
for $X=\mathcal{F},\mathcal{R}$. Next, the integral in \cref{eq:continuous_dot} is approximated by the trapezoidal rule,
\begin{align}\label{eq:dot_discrete}
\dot{U}_N&=-iL_sU_N
+\Delta t\left[\frac{1}{2}\mathcal{K}_N+\sum_{m=1}^{N-1}\mathcal{K}_{m}U_{N-m}+\frac{1}{2}\mathcal{K}_0U_N\right]\notag\\
&+\Delta t^3\sum_{m=1}^{N}D^\mathcal{K}_m,
\end{align}
where  
\begin{align}
D^\mathcal{K}_m&\sim\frac{d^2}{d\tau^2}\left[\mathcal{K}(\tau)U(t-\tau)\right]_{\tau=m\Delta t}.
\end{align}
Similarly to \cref{eq:U1}, the expansion of $U_{N+1}$ at $t=N\Delta t$ follows 
\begin{align}\label{eq:Udiscrete}
U_{N+1}
&=U_N+\Delta t\dot{U}_N+\frac{\Delta t^2}{2}\ddot{U}_N+\frac{\Delta t^3}{6}\dddot{U}_N+\mathcal O(\Delta t^4)\notag\\
&=LU_N\notag
+\Delta t^2\left[\sum_{m=1}^{N}\mathcal{K}_{m}U_{N-m}+K_0U_N\right]\notag\\
&+\frac{\Delta t^3}{2}\sum_{m=1}^{N}\mathcal{F}_{m}U_{N-m}+\mathcal{O}\left(\Delta t^4\sum_{m=1}^N(D^\mathcal{K}_m+D^\mathcal{F}_m)\right)\notag\\
&+\frac{\Delta t^4}{6}\sum_{m=0}^{N-1}\mathcal{R}_{N-m}U_m+\mathcal{O}\left(\Delta t^5\sum_{m=1}^ND^\mathcal{R}_m\right).
\end{align}
Finally, by comparing \cref{eq:nz_discrete} and \cref{eq:Udiscrete}, 
we observe
\begin{align}\label{eq:KN}
K_N&=\mathcal{K}_N+\frac{\Delta t}{2}\mathcal{F}_{N}+\mathcal O\left(\Delta t^2(D_N^\mathcal{K}+D_N^\mathcal{F})\right).
\end{align}

\cref{eq:K0,eq:KN} shows the relationship between $K_N$ and $\mathcal{K}_N$ as follows: 
Up to $O(\Delta t)$, $K_N$ agrees with $\mathcal{K}_N$ for $N>0$, and $K_0=(-L_s^2+\mathcal{K}_0)/2$; this way of relating $\mathcal{K}_N$ and $K_N$ will henceforth be called TTM(1). Furthermore, given exact $\dddot{U}_0$ and $\mathcal{F}_n$, $\mathcal{K}_N$ and $K_N$ can be computed from each other with $O(\Delta t^2)$ error for all $N\geq0$ using \cref{eq:K0,eq:KN}; we refer to this scheme as TTM(2). Note that the TTM(2) scheme requires accurate computation of $\dddot{U}_0$ and $\mathcal{F}_n$. In particular, the integral in \cref{eq:F} needs to be computed on dense grid, making the TTM(2) scheme much more costly. In any case, our analysis reveals that the simple identification of $K_N$ with $\mathcal{K}_N$ (e.g. Eq. (5) in Ref.~\onlinecite{PhysRevLett.112.110401} and Eq. (12) in Ref.~\onlinecite{Pollock2018tomographically}) are valid for $N>0$, and needs to be corrected as \cref{eq:K0} for $N=0$. 

We stress that \cref{eq:K0,eq:KN} only applies when one computes $\mathcal{K}_N$ from exact $U_N$ at discrete times.  From this perspective, the TTM(1) and TTM(2) schemes can be considered as alternative ways to extract continuous time memory kernels $\{\mathcal{K}_m\}$ (at discrete time points) from dynamics simulation, with controllable error. In particular, $\{U_m\}$ and $\{\mathcal{K}_m\}$ are only computed on coarse time grid in the TTM(1) scheme, whereas existing approaches typically involves integral equations hence requires $\{U_m\}$ and $\{\mathcal{K}_n\}$ with fine time resolution~\cite{10.1063/1.5047446,10.1063/1.4948612,10.1063/5.0078040,10.1063/1.2218342,10.1063/1.4948408,10.1063/1.4975388,doi:10.1021/acs.jpcb.1c05719,10.1063/1674-0068/cjcp2109157}.

Another use case is propagating $U_N$ according to \cref{eq:nz_discrete}, using $K_N$ computed from exact $\mathcal{K}_N$ as in \cref{eq:K0,eq:KN}. In this case, the above analysis does not quantify how the discretization error propagates. In the next section, we will present numerical results for the various schemes in both scenarios: computing $\mathcal{K}_N$ from the exact $U_N$, and computing $U_N$ from the exact $\mathcal{K}_N$.

In Ref.~\onlinecite{doi:10.1021/acs.jctc.5c00396}, an alternative discretization scheme was proposed using the midpoint derivative and midpoint integrals to improve accuracy. In this scheme (hereafter denoted as MPD/I), the discrete time memory kernels $T_N$ are computed as 
\begin{align}\label{eq:mpdi_TU}
\frac{1}{2}T_1&=U_1-\frac{1}{2}G_1\\
\frac{1}{2}T_N&=U_N-\sum_{m=1}^{N-1}T_{N-m}U_{m},\quad N>1,
\end{align}
and $\mathcal{K}_N$ computed from $T_N$ as
\begin{align}\label{eq:mpdi_TK}
&\Delta t^2G_{1/2}\mathcal{K}_{1/2}=T_1-G_1\\
&\Delta t^2G_{1/2}\mathcal{K}_{N-1/2}=T_N,\quad N>1
\end{align}
where $G_1=e^{-i\Delta tL_s}$ and $G_{1/2}=e^{-i\Delta tL_s/2}$, and $\mathcal{K}_N=(\mathcal{K}_{N-1/2}+\mathcal{K}_{N+1/2})/2$~\cite{note}. 
We will also compare the TTM(1) and TTM(2) schemes with MPD/I in the numerical results.

Finally, we emphasize that the original TTM method~\cite{PhysRevLett.112.110401} uses only discrete time memory kernel $\{K_m\}$ to propagate dynamics, which in turn are computed from short time dynamical channels $\{U_m\}$, both according to \cref{eq:nz_discrete}. In particular, it does not require a continuous-time memory kernel $\{\mathcal{K}_n\}$ and is free of any additional discretization error beyond that of $\{U_m\}$. Hence, given exact short time $\{U_m\}$, the only error in the original TTM method comes from memory time truncation, i.e., assuming $K_n=0$ for all $n>n_T$ in \cref{eq:nz_discrete}, where $n_T$ is some memory truncation time. 

\section{Numerical results and discussion}\label{sec:numerics}

\begin{figure*}[htbp]
    \centering
    \begin{subfigure}{0.49\linewidth}
        \centering
        \includegraphics[width=1\linewidth]{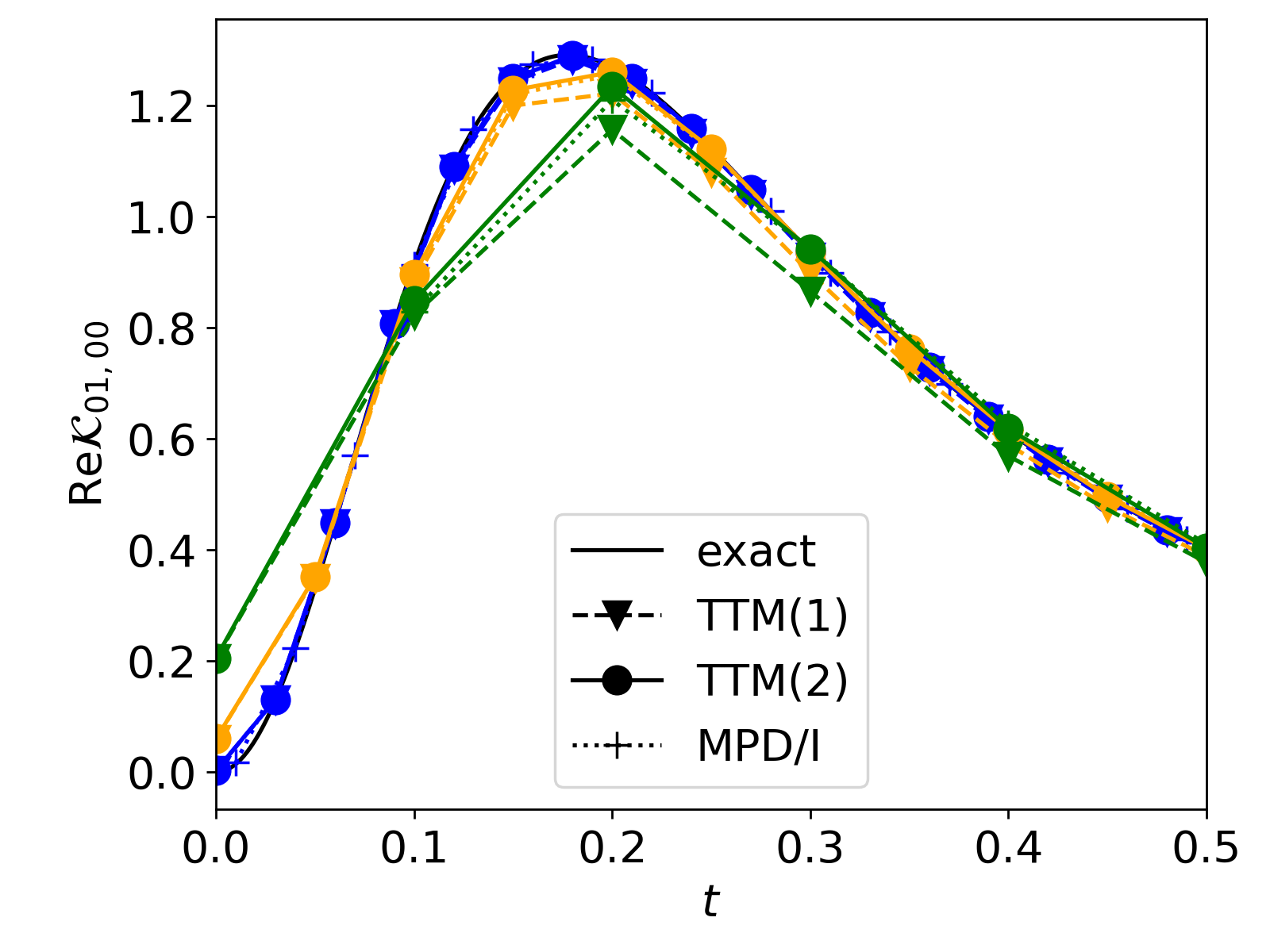}
        \caption{}
        \label{fig:T10_real}
    \end{subfigure}
    \centering
    \begin{subfigure}{0.49\linewidth}
        \centering
\includegraphics[width=1\linewidth]{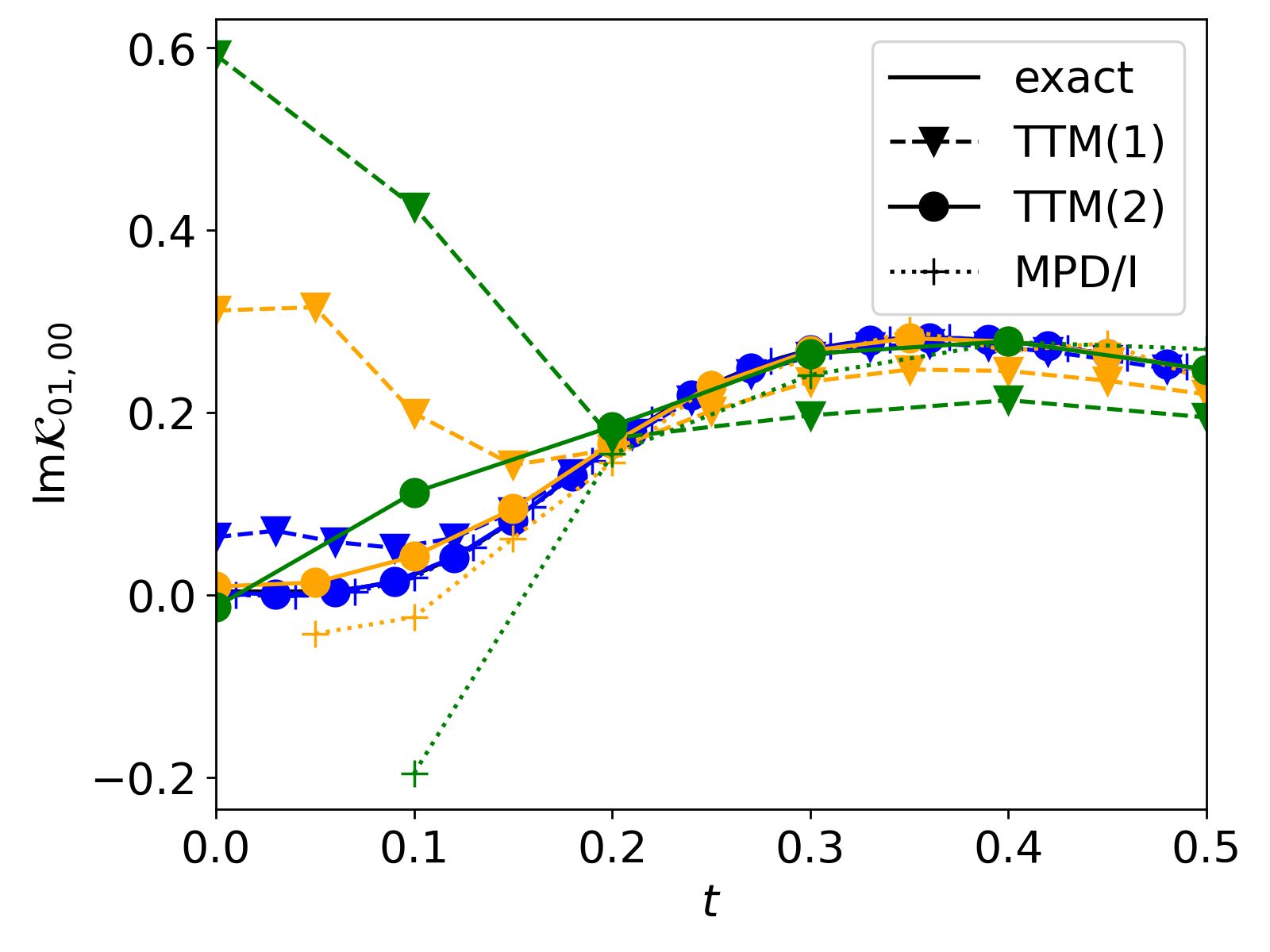}
        \caption{}
        \label{fig:T10_imag}
    \end{subfigure}
    \centering
    \begin{subfigure}{0.49\linewidth}
        \centering
        \includegraphics[width=1\linewidth]{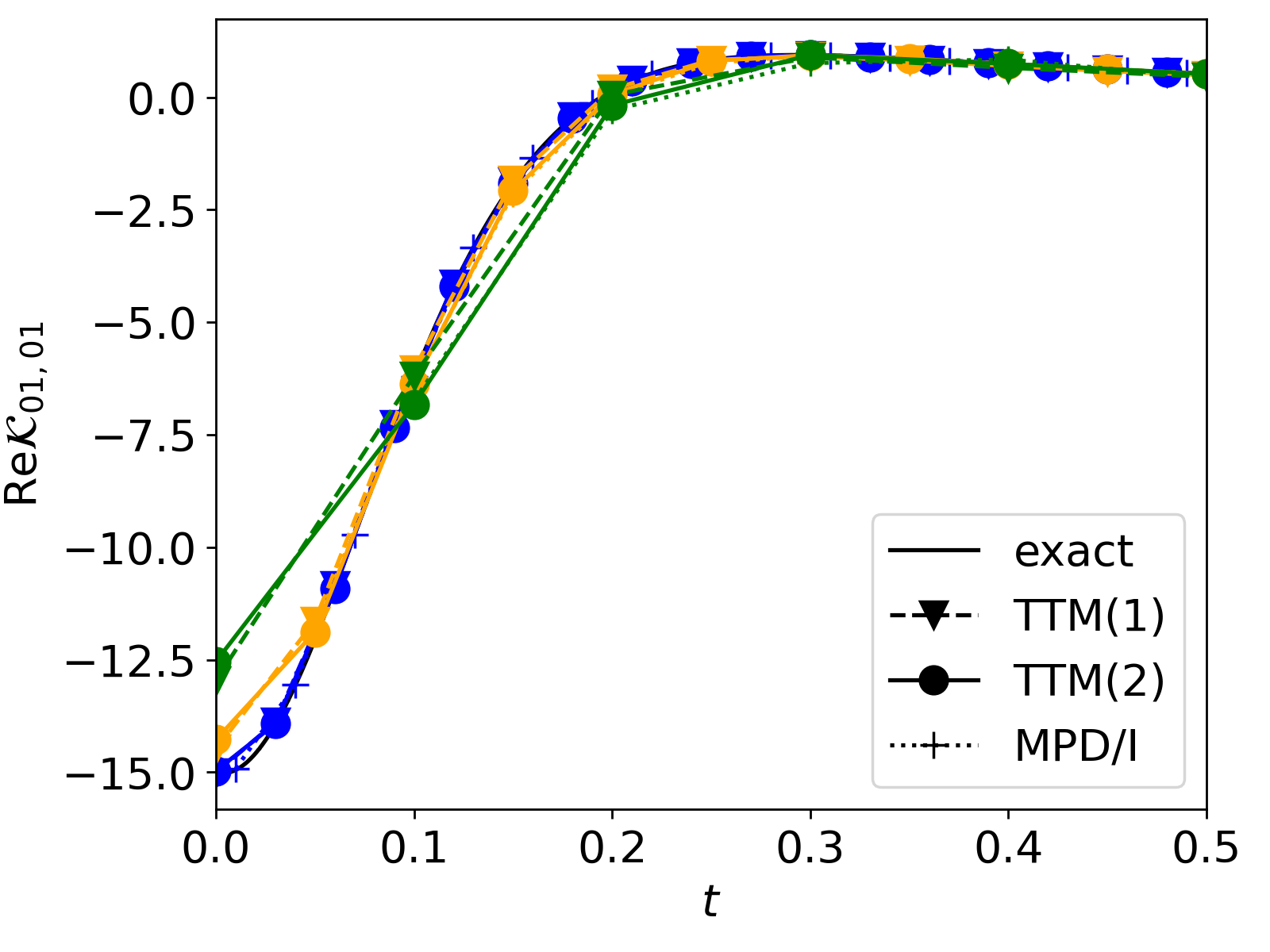}
        \caption{}
        \label{fig:T11_real}
    \end{subfigure}
    \centering
    \begin{subfigure}{0.49\linewidth}
        \centering
        \includegraphics[width=1\linewidth]{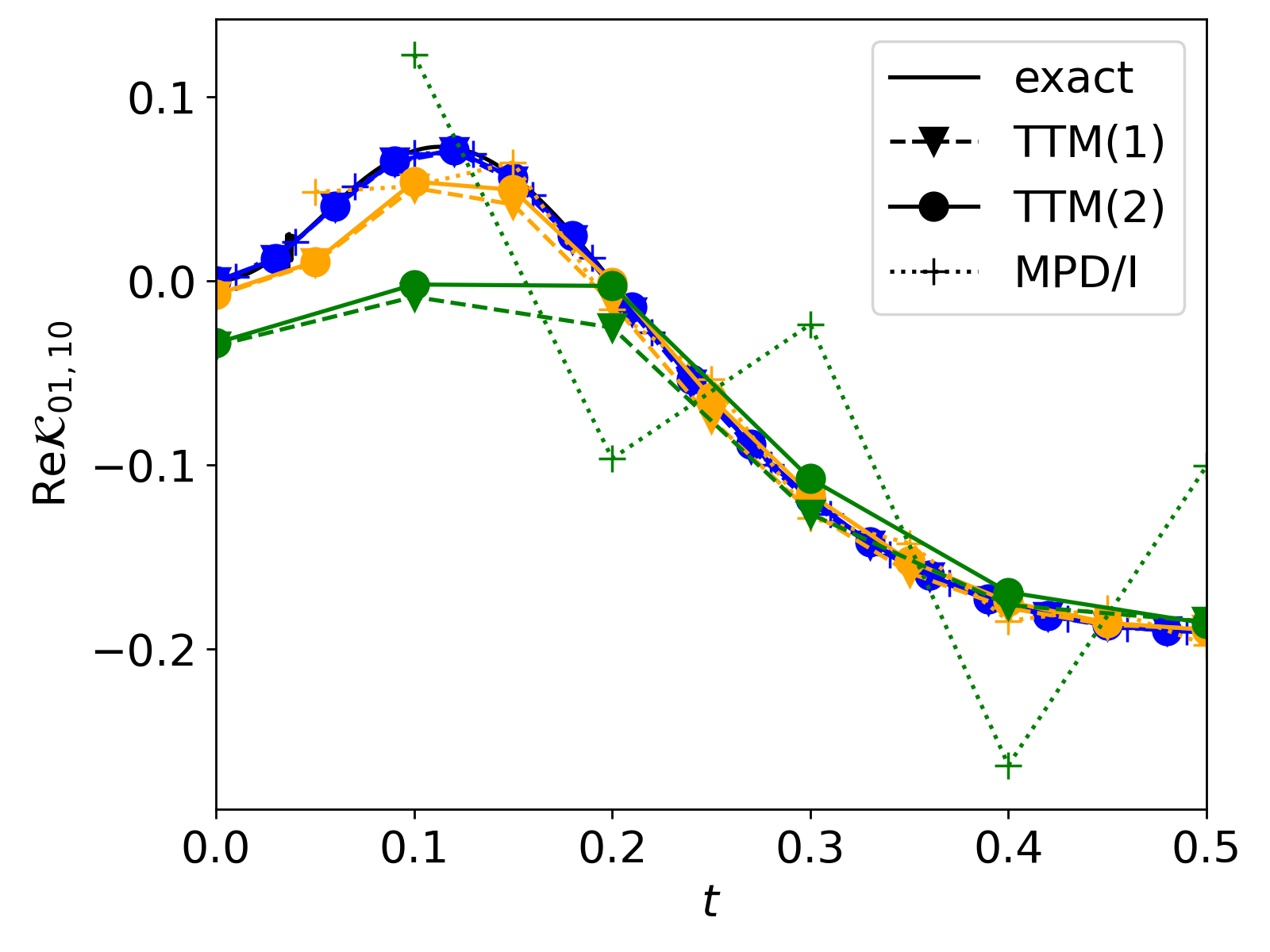}
        \caption{}
        \label{fig:T12_real}
    \end{subfigure}
    \caption{Elements of continuous time memory kernel $\mathcal{K}$ from TTM(1), TTM(2) and MPD/I, with corresponding markers labeled as in the plot. Exact results are shown in the black solid curve without a marker. Blue, orange, and green correspond to $\Delta t=0.01$, 0.05, and 0.1, respectively. 
    }
    \label{fig:kernel}
\end{figure*}

\begin{figure}[htbp]
    \centering
    \begin{subfigure}{\linewidth}
    \centering
    \includegraphics[width=\linewidth]{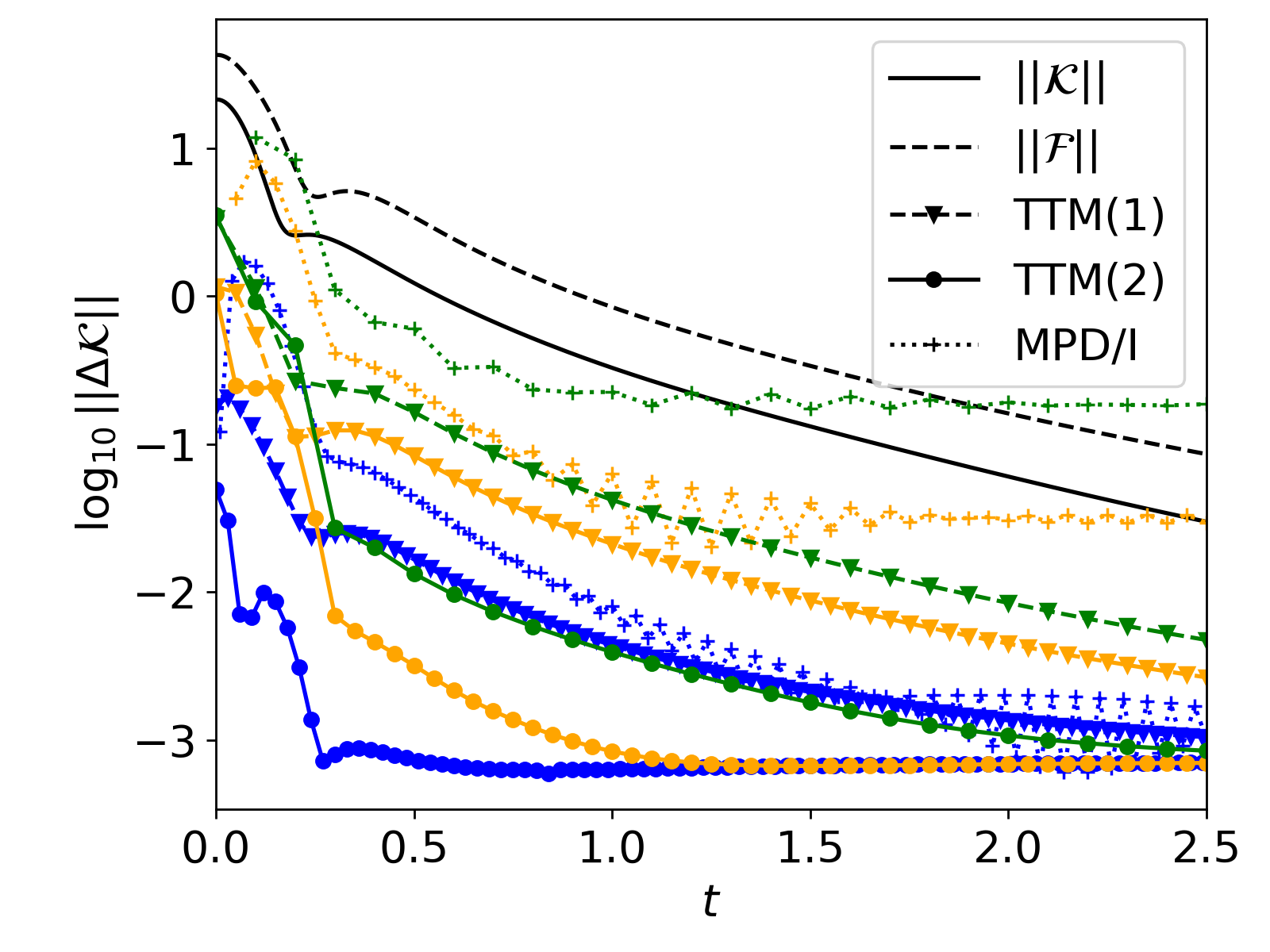}  
    \caption{}
    \label{fig:ker_err_0.0005}
    \end{subfigure}
    
    \centering
    \begin{subfigure}{\linewidth}
    \centering
    \includegraphics[width=\linewidth]{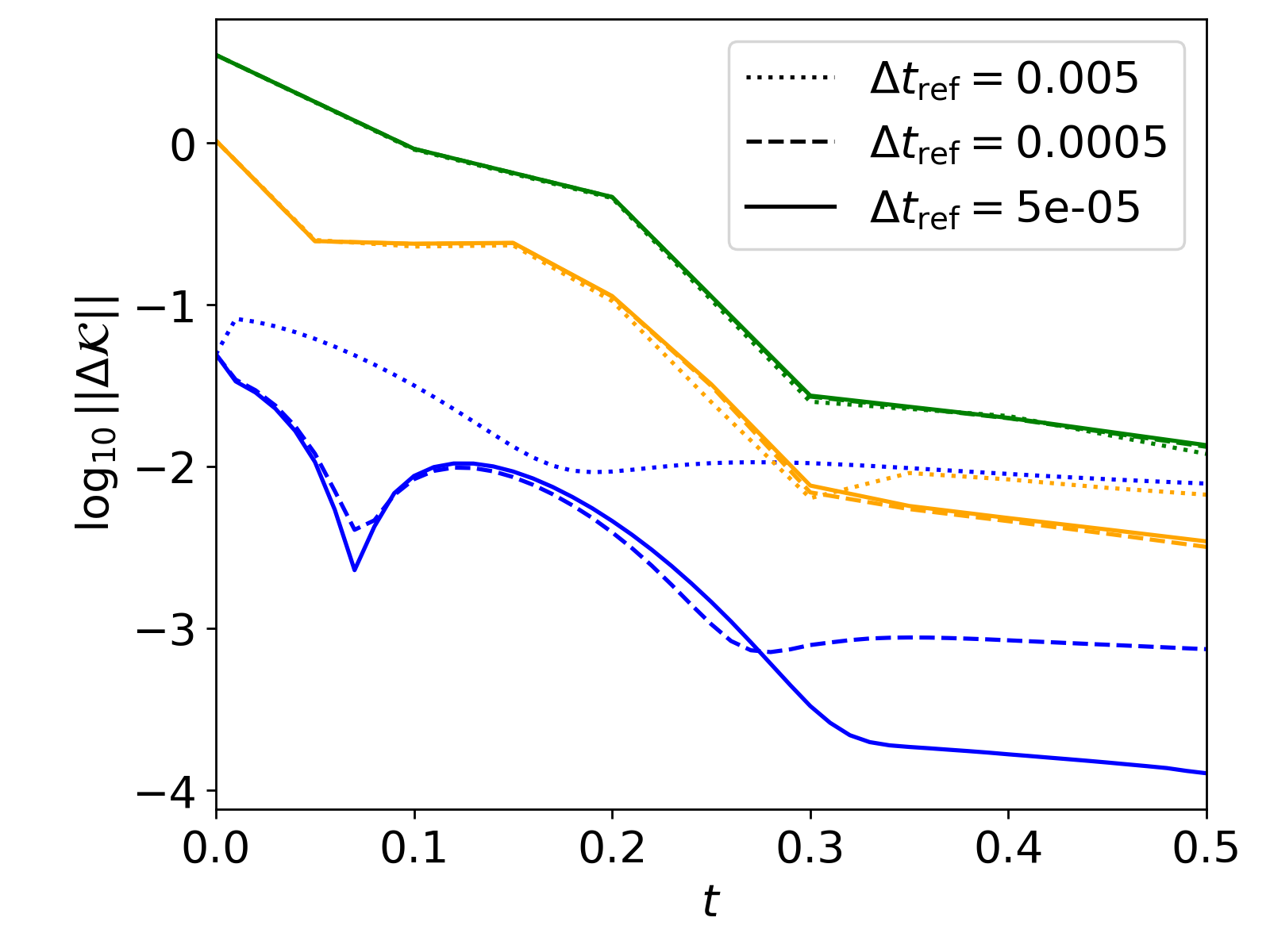}  
    \caption{}
    \label{fig:ker_err_5e-5}
    \end{subfigure}
    \caption{Error of continuous time memory kernel $\mathcal{K}$. (a) Results from TTM(2), TTM(1), and MPD/I schemes are labeled as Fig.~\ref{fig:kernel}. Reference $\log_{10}||\mathcal{K}_N||$ and $\log_{10}||\mathcal{F}_N||$ computed with $\Delta t_\mathrm{ref}=0.0005$ are also plotted in black solid and dashed curves, respectively. (b) TTM(2) results, where the reference $\mathcal{K}_N$ and $\mathcal{F}_N$ are computed using increasing $\Delta t_\mathrm{ref}$ as labeled. 
    In both panels, blue, orange, and green again correspond to $\Delta t=0.01$, 0.05, and 0.1, respectively.
    }
    \label{fig:kernel_err}
\end{figure}

\begin{figure*}[htbp]
    \centering
    \begin{subfigure}{0.329\linewidth}
        \centering
        \includegraphics[width=1\linewidth]{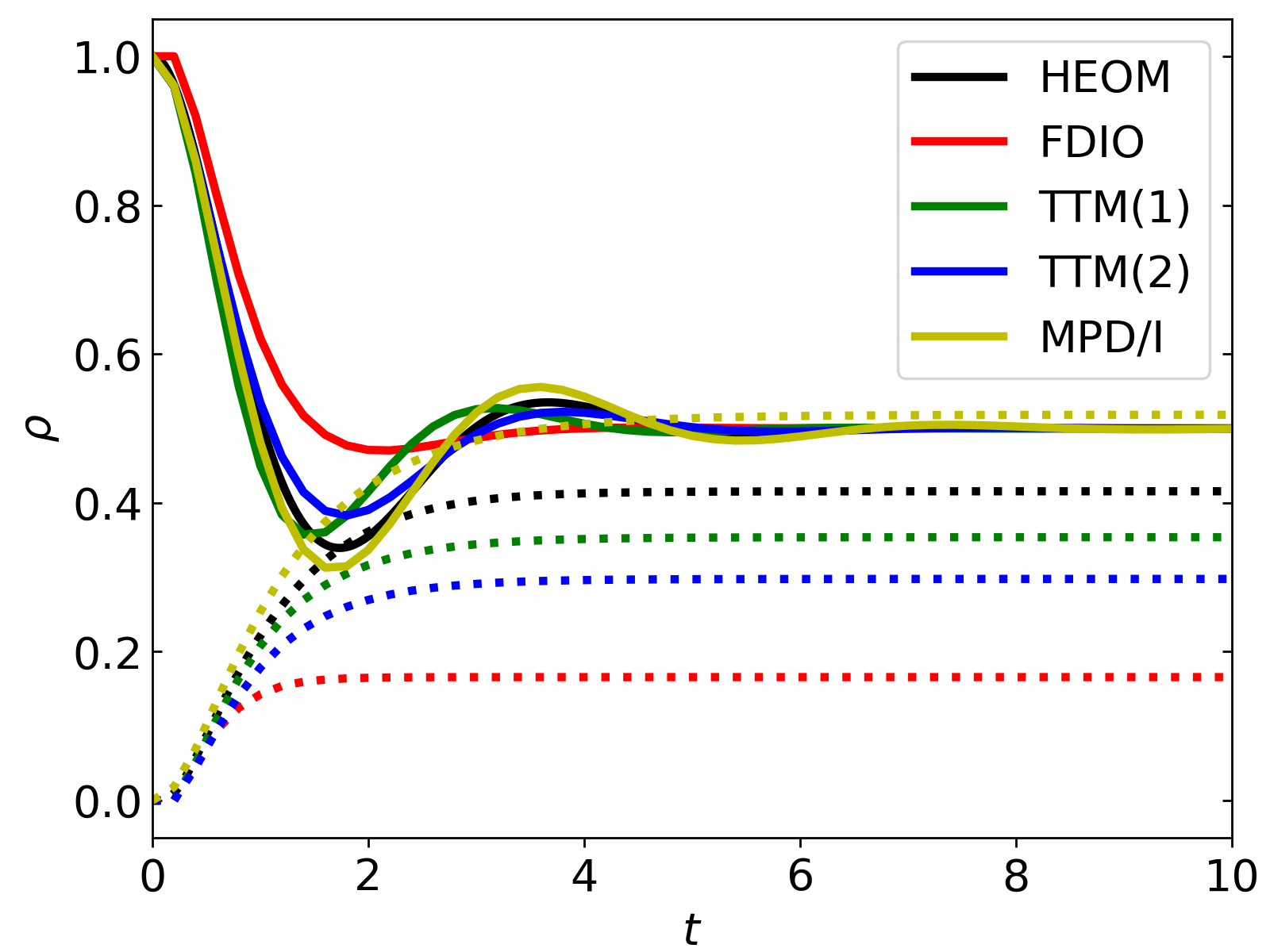}
        \caption{}
        \label{fig:rho_0.2}
    \end{subfigure}
    \centering
    \begin{subfigure}{0.329\linewidth}
        \centering
\includegraphics[width=1\linewidth]{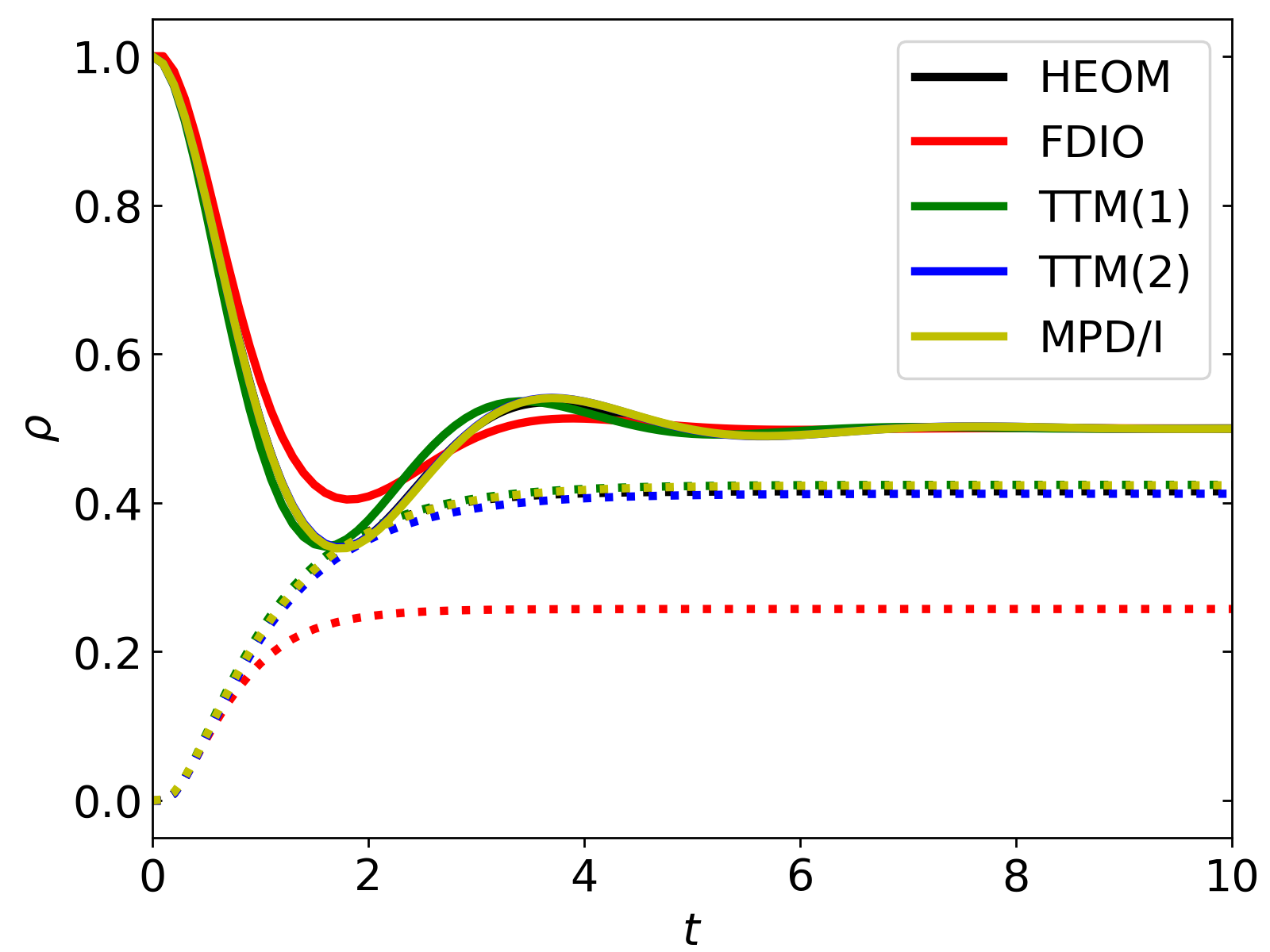}
        \caption{}
        \label{fig:rho_0.1}
    \end{subfigure}
    \centering
    \begin{subfigure}{0.329\linewidth}
        \centering
        \includegraphics[width=1\linewidth]{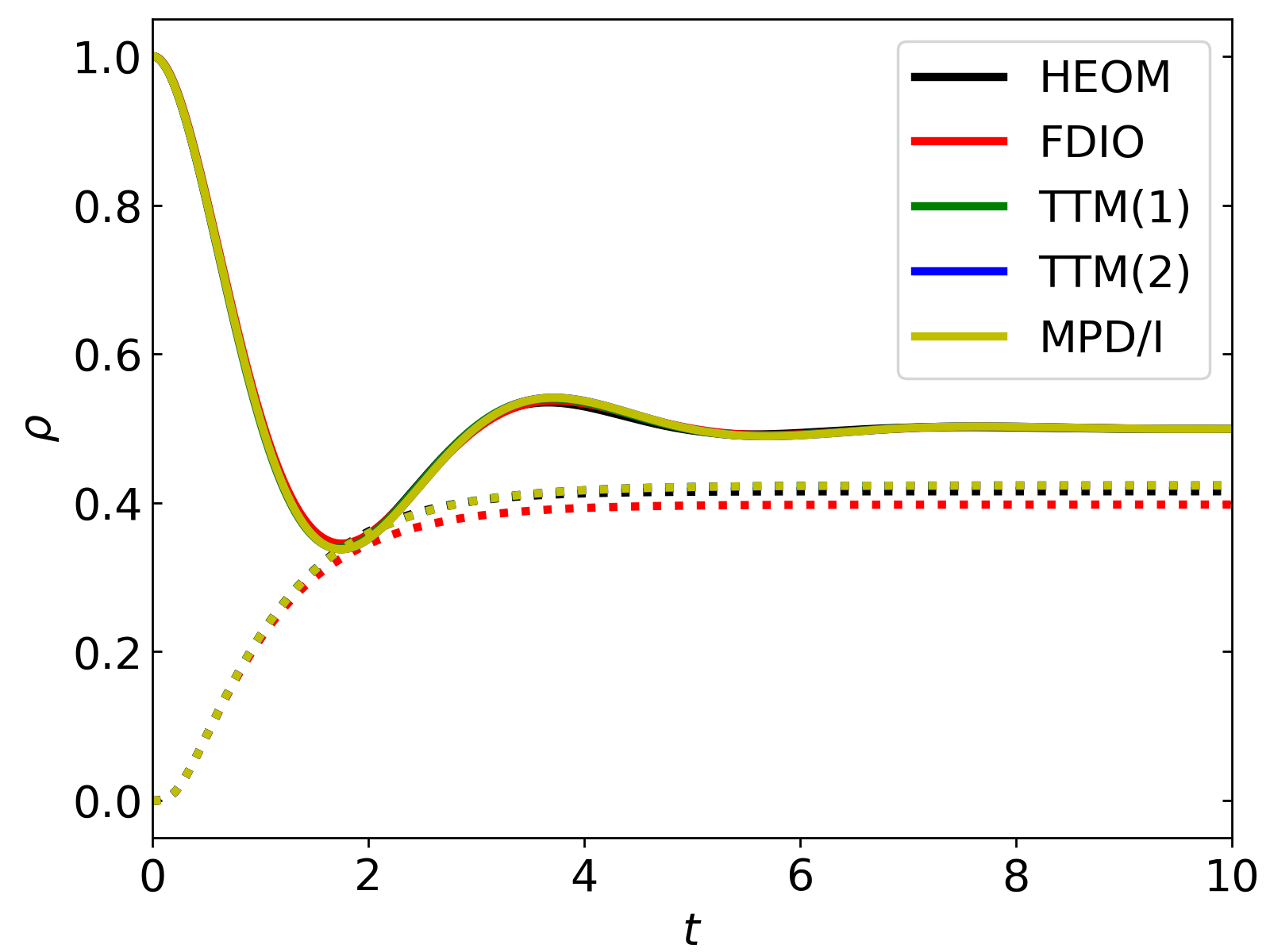}
        \caption{}
        \label{fig:rho_0.01}
    \end{subfigure}
    \caption{RDM elements Re$\rho_{00}$ (solid) and Re$\rho_{01}$ (dotted) as a function of time. In each panel, black curves are computed from HEOM with $\Delta t=0.0005$. FDIO, TTM(1), TTM(2), and MPI/D are plotted in corresponding colors with (a) $\Delta t=0.2$, (b) $\Delta t=0.1$, and (c) $\Delta t=0.01$. }
    \label{fig:rho}
\end{figure*}

\begin{figure*}[htbp]
    \centering
    \begin{subfigure}{0.329\linewidth}
        \centering
\includegraphics[width=1\linewidth]{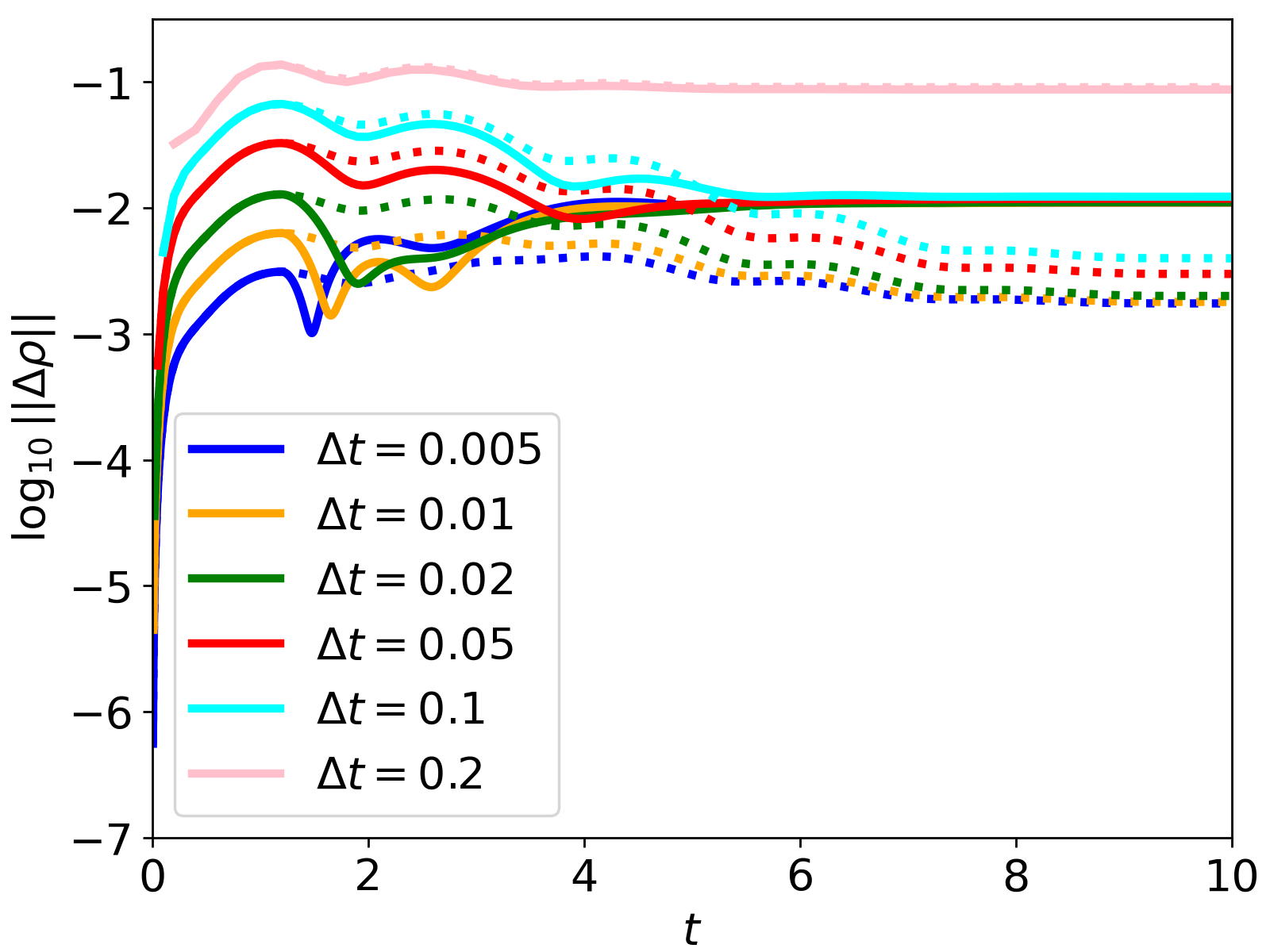}
        \caption{}
        \label{fig:rho_TTM_K}
    \end{subfigure}
    \centering
    \begin{subfigure}{0.329\linewidth}
        \centering
\includegraphics[width=1\linewidth]{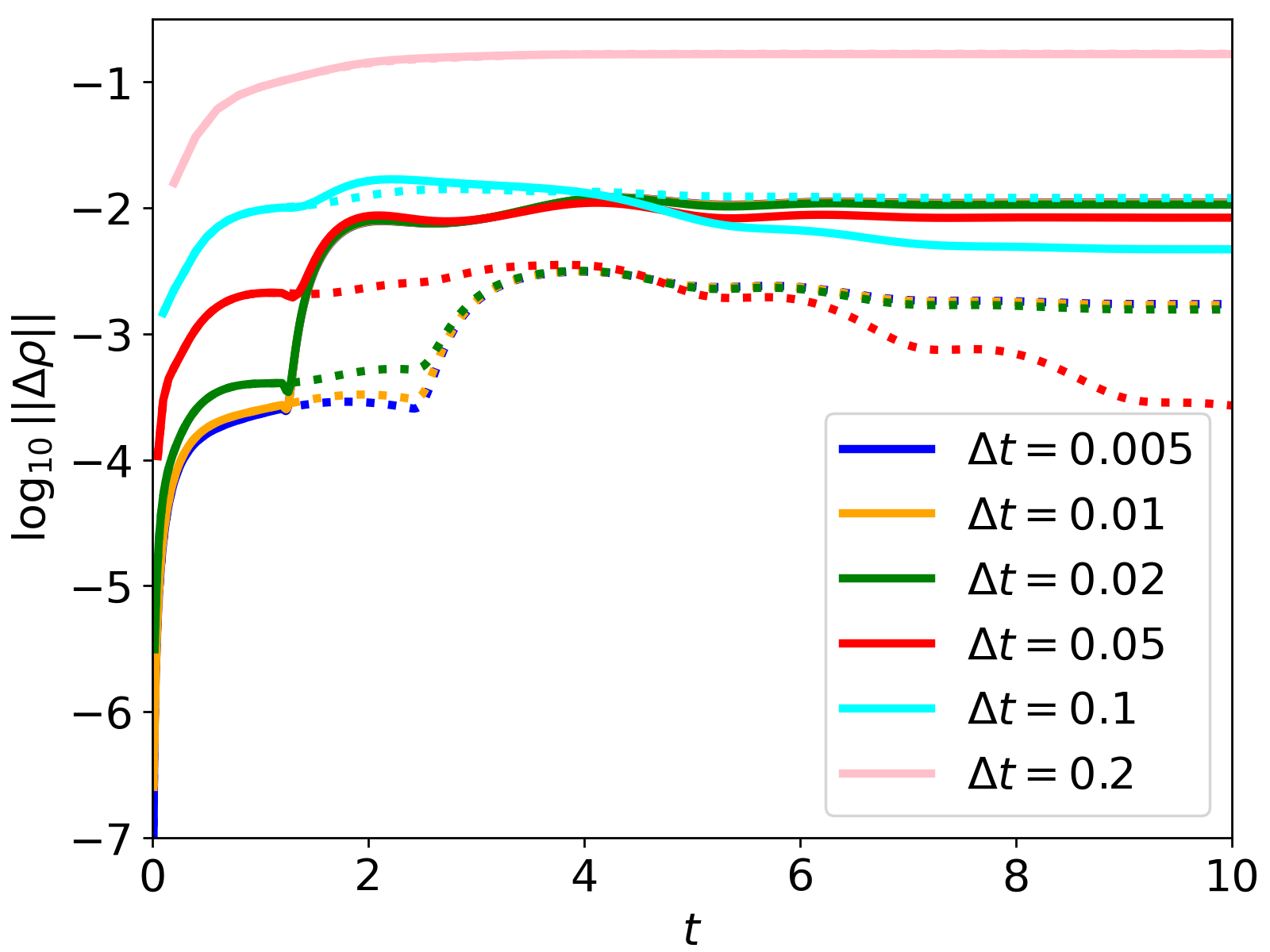}
        \caption{}
        \label{fig:rho_TTM_K2}
    \end{subfigure}
    \centering
    \begin{subfigure}{0.329\linewidth}
        \centering
        \includegraphics[width=1\linewidth]{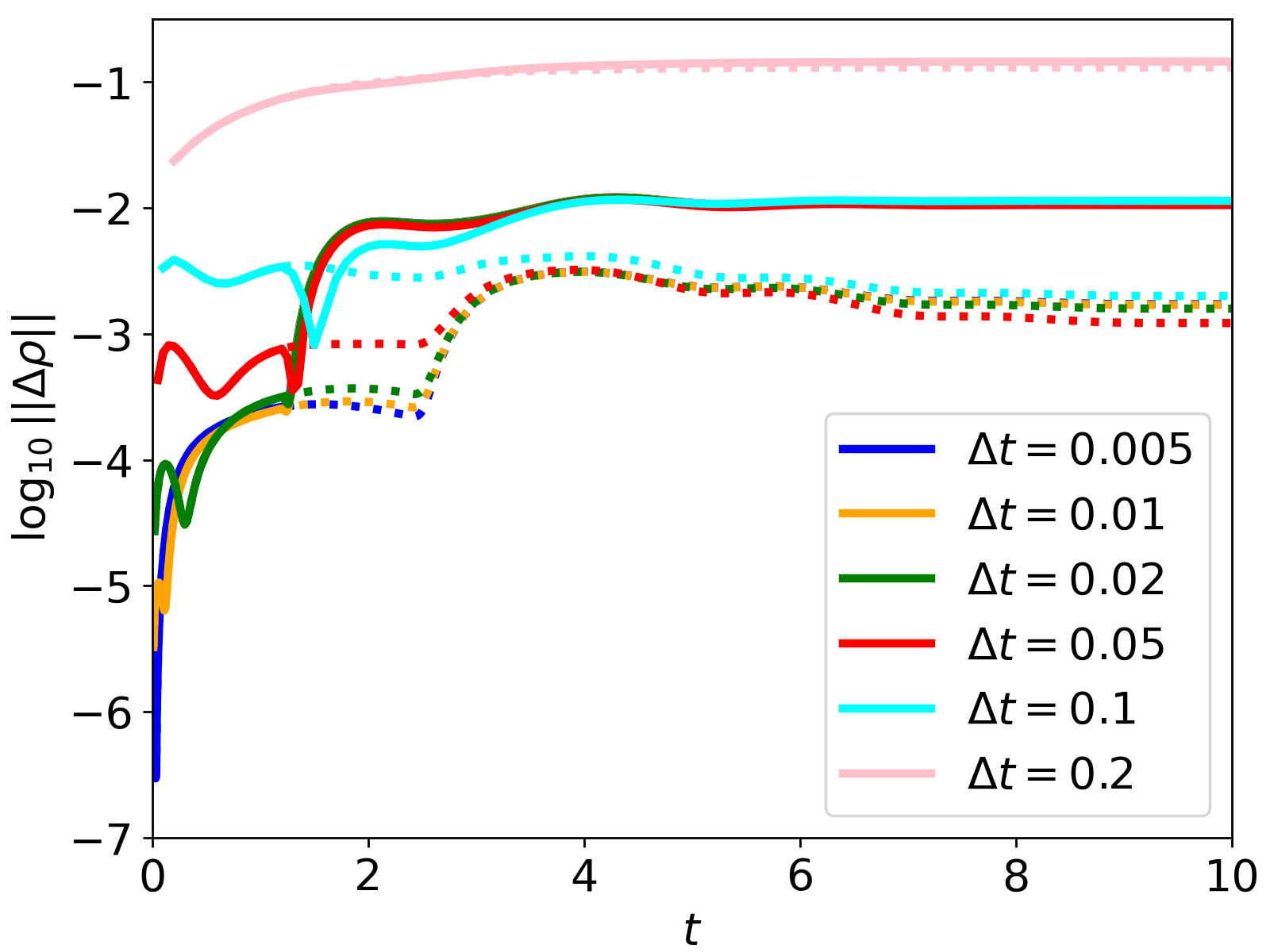}
        \caption{}
        \label{fig:rho_MPDI}
    \end{subfigure}
    \caption{$\log_{10}(||\rho-\rho_{\text{HEOM}}||)$ as a function of time computed from (a) TTM(1), (b) TTM(2), (c) MPD/I. In each panel, solid and dotted curves are computed with memory cutoffs, $t_\text{mem}=1.2$ and 2.4, respectively. Different time step sizes $\Delta t$ are plotted in corresponding colors. }
    \label{fig:drho}
\end{figure*}

We now move to numerical illustrations of our analysis. We will consider the archetypal open-quantum system Hamiltonian, the spin-boson model Hamiltonian, given as
\begin{align}
\hat{H}&=\hat{H}_s+\hat{H}_b\\
\hat{H}_s&=\epsilon\sigma_z+\Omega\sigma_x\\
\hat{H}_b&=\sum_j\frac{\hat{p}_j^2}{2m}+\frac{1}{m}\omega_j^2\hat{q}_j^2-c_j\hat{q}_j\sigma_z,
\end{align}
where $\sigma_z$, $\sigma_x$ are Pauli matrices, $\omega_j$ and $c_j$ are bath frequencies and system-bath coupling coefficients collectively specified through the spectral density function
\begin{align}
J(\omega)=\frac{\pi}{2}\xi\left(\frac{\omega^s}{\omega_c^{s-1}}\right)e^{-\omega/\omega_c},
\end{align}
where $\omega_c$ is the cutoff frequency and $\xi$ is the Kondo parameter.
For comparison, we use the same parameters as in Ref.~\onlinecite{doi:10.1021/acs.jctc.5c00396}: $\epsilon=0$, $\Omega=-1$, $s=1$, $\omega_c=5$, $\xi=0.3$ and inverse temperature $\beta=5$. 

Furthermore, in the following, the exact $U_N$ are assembled from simulation of independent trajectories using the hierarchical equation of motion (HEOM)~\cite{tanimura1989time,10.1063/5.0011599,PhysRevLett.129.230601,10.1063/1.5026753,mangaud2023survey} combined with a tensor network representation of the Auxiliary Density Operators (ADOs) \cite{10.1063/5.0050720,Mangaud2023,lindoy2025pyttnopensourcetoolbox} (further details of the implementation are provided in Ref.~\onlinecite{ivander2024unified}). 
In each of these calculations, the bath correlation function was fit to a sum-of-exponential form
\begin{equation}
C(t) = \sum_{k=1}^K \alpha_k e^{-\nu_k t},
\end{equation}
using the Adaptive Antoulas-Anderson (AAA) algorithm~\cite{doi:10.1137/16M1106122} to construct a rational function approximation to the bath spectral function~\cite{PhysRevLett.129.230601}.  For the ohmic spin-boson model considered here, a total of $K=7$ terms were sufficient to accurately capture the bath correlation function over the time scales considered, corresponding to a hierarchy of ADOs containing 14 modes.  A direct product basis was used for the hierarchy of ADOs, with each mode, $k$, truncated at a depth 
\begin{equation}
L_k = \mathrm{min}\left(L_{\mathrm{max}}, \mathrm{max}\left(L_{\mathrm{min}},  \frac{L_{\mathrm{max}}\nu_{\mathrm{min}}}{\mathrm{Re}(\nu_k)}\right)\right),
\end{equation}
where $L_\mathrm{max}=20$ is the maximum allowed depth for a given mode, $L_\mathrm{min}=8$ is the minimum depth to use for any given mode, and $\nu_{\mathrm{min}}$ is the smallest decay rate in the sum-of-exponential decomposition of the bath correlation function.
Then $\mathcal{K}_N$ are the numerically converged solutions of the integro-differential equation as described in Ref.~\onlinecite{10.1063/1.5047446}, and $\mathcal{F}_N$ are computed from \cref{eq:F}. 
To compute the numerically ``exact'' $\mathcal{K}_N$, $\mathcal{F}_N$, and $\dddot{U}_0$, we use a small time step size, $\Delta t_\mathrm{ref}=0.0005$, unless specified otherwise. 

We first consider computing $\mathcal{K}_N$ from exact $U_N$. 
Fig.~\ref{fig:kernel} plots the elements of $\mathcal{K}$ (chosen to be the same as in Fig. 3 and Fig. 4 of Ref.~\onlinecite{doi:10.1021/acs.jctc.5c00396}) computed from discrete memory kernels using various schemes (labeled by different markers and line styles) discussed above. Exact $\mathcal{K}$ are shown in black solid curves without markers. We first note that at $N=0$, $\mathcal{K}_0$ computed from both TTM schemes converges to the exact result as $\Delta t\to0$, suggesting that the ``spurious term'' identified in Ref.~\onlinecite{doi:10.1021/acs.jctc.5c00396} can indeed be corrected by \cref{eq:K0}.
Furthermore, MPD/I is not uniformly more accurate than TTM(1) with the same $\Delta t$: For real and imaginary part of $\mathcal{K}_{01,00}$ (Fig.~\ref{fig:T10_real} and Fig.~\ref{fig:T10_imag}), we reproduce the results in Fig. 4 of Ref.~\onlinecite{doi:10.1021/acs.jctc.5c00396}, where MPD/I is indeed more accurate. 
Yet for Re$\mathcal{K}_{01,01}$ and Re$\mathcal{K}_{01,01}$ (Fig.~\ref{fig:T11_real} and Fig.~\ref{fig:T12_real}), TTM(1) is as accurate as, if not more accurate than, MPD/I. 

Fig.~\ref{fig:ker_err_0.0005} plots the Frobenius norm of error $||\Delta\mathcal{K}||$ of the various schemes, with the reference $\mathcal{K}$ computed with $\Delta t_\mathrm{ref}=0.0005$. 
We first note that, the MPD/I error $||\Delta \mathcal{K}_N||$ approaches an $N$-independent constant in the large $N$ limit, where the constant scales as $\mathcal{O}(\Delta t^2)$. In particular, for large $N$, the MPD/I scheme leads to an error $\Delta \mathcal{K}_N$ which exceeds the value of $\mathcal{K}_N$ itself, and hence cannot predict $\mathcal{K}_N$ correctly.
For TTM(1), \cref{eq:KN} predicts an error $||\Delta\mathcal{K}_N||=\mathcal{O}(||\mathcal{F}_N||\Delta t)$, which is numerically confirmed in Fig.~\ref{fig:ker_err_0.0005} by the decay of $||\Delta \mathcal{K}||$ which parallels that of $||\mathcal{F_N}||$. 
For TTM(2), \cref{eq:KN} predicts an error $||\Delta \mathcal{K}_N||=O\left(\Delta t^2(||D^\mathcal{K}_N||+||D^\mathcal{F}_N||)\right)$, where $D^\mathcal{K}_N$ and $D^\mathcal{F}_N$ decays as derivatives of $\mathcal{K}$ and $\mathcal{F}$. 
This is numerically demonstrated in Fig.~\ref{fig:ker_err_0.0005} by the initial fast decay until $||\Delta\mathcal{K}||\approx0.001$. 
Furthermore, the saturation at $||\Delta\mathcal{K}||\approx0.001$ is due to numerical error in computing reference $\mathcal{K}$ and $\mathcal{F}$ with $\Delta t_\mathrm{ref}=0.0005$. 
This is confirmed in Fig.~\ref{fig:ker_err_5e-5}, where the reference $\mathcal{K}$ and $\mathcal{F}$ are computed with three different values of $\Delta t_\mathrm{ref}$. 
In each case, the TTM(2) $||\Delta\mathcal{K}_N||$ (using different $\Delta t$) decrease with $N$, until saturates at around $2\times\Delta t_\mathrm{ref}$.%

Next, we consider simulating dynamics using the exact $\mathcal{K}_N$. Fig.~\ref{fig:rho} plots the long-time system density matrix (RDM) elements Re$\rho_{00}$ (solid) and Re$\rho_{01}$ (dotted), with initial state $\rho(0)=|0\rangle\langle0|$, and with panel~\ref{fig:rho_0.2},~\ref{fig:rho_0.1} and~\ref{fig:rho_0.01} corresponds to $\Delta t=0.2$, 0.1, and 0.01 respectively. In each panel, black curves are the exact results computed from HEOM with time step size $\Delta t=0.0005$. Different discretization schemes are plotted in corresponding colors, where the same memory cutoff $t_{\text{mem}}=1.2$ is used as suggested in Ref.~\onlinecite{doi:10.1021/acs.jctc.5c00396}. In addition, we also plot the result corresponding to simple identification $K_N=\mathcal{K}_N$ for all $N\geq0$~\cite{PhysRevLett.112.110401,Pollock2018tomographically}, also called ``forward-difference approximation with integral overestimation'' (FDIO) in Ref.~\onlinecite{doi:10.1021/acs.jctc.5c00396}. The FDIO results we obtained is manifestly different from that of Ref.~\onlinecite{doi:10.1021/acs.jctc.5c00396} (see Fig. 5 therein): (i) our FDIO results shows discretization error as soon as time step $N=1$, whereas those of Ref.~\onlinecite{doi:10.1021/acs.jctc.5c00396} seems numerically exact for $t<t_\mathrm{mem}$; (ii) FDIO dynamics in Ref.~\onlinecite{doi:10.1021/acs.jctc.5c00396} shows a kink at $t_\mathrm{mem}$, where as our FDIO dynamics are smooth. However, our FDIO result agrees with that of Ref.~\onlinecite{doi:10.1021/acs.jctc.5c00396} in the long time time limit. 

Fig.~\ref{fig:drho} plots the $\log_{10}||\rho-\rho_{\text{HEOM}}||$ for the schemes used in Fig.~\ref{fig:rho} with different $\Delta t$, where panel (a),(b),(c) corresponds to TTM(1), TTM(2) and MPD/I, respectively. For each approximation, solid curves correspond to memory cutoff $t_\text{mem}=1.2$ and dotted curves correspond to $t_{\text{mem}}=2.4$.

Combining observations from Fig.~\ref{fig:rho} and Fig.~\ref{fig:drho}, we make the following observations:
(i) Below the memory cutoff time (i.e., $t\leq t_\text{mem}$), the error convergence in $O(\Delta t)$ is slowest for TTM(1), and comparable between TTM(2) and MPDI. (ii) Beyond the memory cutoff time, the errors in TTM(1), TTM(2), and MPD/I are all dominated by memory truncation and less affected by discretization error. These observations suggest that, if one can obtain an accurate continuous time memory kernel $\mathcal{K}_N$ without access to dynamical channels $U_N$, then MPD/I could yield accurate short time dynamics than TTM(1) and TTM(2).

\section{Conclusion}

This Communication discusses the transfer tensor method (TTM) as a particular discretization of the Nakajima-Zwanzig quantum master equation (NZ-QME). We demonstrated the relationship between the discrete-time memory kernel $K_N$ in TTM and the continuous-time NZ-QME kernel $\mathcal{K}(N\Delta t)$, evaluated at discrete time points, up to an additive $\mathcal {O} (\Delta t^2)$ error. Specifically, we showed that, for $N=0$, $K_0=(-L_s^2+\mathcal{K}_0)/2+\Delta t\dddot{U}_0/6+\mathcal O(\Delta t^2)$ and for $N>0$, $K_N=\mathcal{K}_N+\Delta t\mathcal{F}_N/2+\mathcal O(\Delta t^2)$, where $\mathcal{F}_N$ can be exactly computed from continuous $\mathcal{K}(t)$. The truncation at $\mathcal{O}(\Delta t)$ versus $\mathcal{O}(\Delta t^2)$ leads to schemes TTM(1) and TTM(2), respectively. 
In both cases, $N=0$ can be consistently treated to yield the same time-step error as the other time points, offering a resolution to the issues discussed in Ref.~\citenum{doi:10.1021/acs.jctc.5c00396}. The relationship was verified via numerical simulations using the spin-boson model. We observed that the elements of $\mathcal{K}_N$ computed from discrete-time exact dynamical channels, $U_N$, converge to the exact result as $\Delta t\to0$. 
Furthermore, $K_N$ computed from exact $\mathcal{K}_N$ leads to accurate discrete-time dynamics for sufficiently small $\Delta t$ ($\Delta t\leq0.1$). 

As discussed in Ref.~\onlinecite{doi:10.1021/acs.jctc.5c00396}, the discretization of NZ-QME is not unique. Alternatives, such as the midpoint derivative/midpoint integral (MPD/I) scheme proposed in Ref.~\onlinecite{doi:10.1021/acs.jctc.5c00396}, may offer practical advantages. In the spin-boson model example, MPD/I is found to be more efficient in generating accurate short time dynamics from exact $\mathcal{K}_N$, while the TTM discretizations provide $\mathcal{K}_N$ from discrete-time exact dynamical channels with better convergence properties as $\Delta t\to0$.

\section*{Data availability}
Data generated in this study is available on GitHub (\url{https://github.com/JoonhoLee-Group/GQME_discretization_error.git}).

\section*{Code availability}
Simulation codes used in this study are available on GitHub (\url{https://github.com/JoonhoLee-Group/GQME_discretization_error.git}).

\section*{Acknowledgements}
R.P. and J.L. were supported by the U.S. Department of Energy, Office of Science, Accelerated Research in Quantum Computing Centers, Quantum Utility through Advanced Computational Quantum Algorithms, grant no. DE-SC0025572. L.P.L. acknowledges the support of the Engineering and Physical Sciences Research Council [grant EP/Y005090/1].

\bibliography{export}

\begin{thebibliography}{45}
\bibitem [{not()}]{note}%
  \BibitemOpen
  \href@noop {} {}\bibinfo {note} {Note that in Table~1 of Ref.
  \citenum{doi:10.1021/acs.jctc.5c00396} there is an additional factor of $1/2$
  from $T$ to $\mathcal{K}$, which we believe is a typo from comparing their
  Eq.~(33) and (36), and this was also numerically confirmed.}\BibitemShut
  {Stop}%
\end{thebibliography}

\end{document}